\documentclass[twocolumn]{jpsj2}

\usepackage{color}
\makeatletter
\def\tr{\mathop{\operator@font tr}\nolimits}
\makeatother
\makeatletter
\def\Tr{\mathop{\operator@font Tr}\nolimits}
\makeatother

\title{%        %You can use \\ for explicit line-break
Quasi-stationary States of Two-Dimensional Electron Plasma \\
Trapped in Magnetic Field%
}

\def\runauthor{Ryo Kawahara and Hiizu Nakanishi}

\author{
 Ryo \textsc{Kawahara}\thanks{E-mail : ryokawa@stat.phys.kyushu-u.ac.jp} 
 and
 Hiizu \textsc{Nakanishi}\thanks{E-mail : naka4scp@mbox.nc.kyushu-u.ac.jp}}

\recdate{\today}

\kword{
non-neutral plasma,
two-dimensional turbulence,
numerical simulation,
long-range force,
nonextensive system,
}

\inst{%         %Affiliation, neglected when [addenda] or [errata]
Department of Physics, Kyushu University 33,
Fukuoka 812-8581
}

\abst{%         
We have performed numerical simulations on a pure electron plasma system
under a strong magnetic field, in order to examine
 quasi-stationary states that the system eventually evolves
into.
  We use ring states as the initial states,
changing the width, and find that the system evolves into a vortex crystal
state from a thinner-ring state while a state with a single-peaked
density distribution is obtained from a thicker-ring initial state.
  For those quasi-stationary states, density
distribution and macroscopic observables are defined on the basis of a
coarse-grained density field.
  We compare our results with experiments and some statistical
theories, which include the Gibbs-Boltzmann statistics, Tsallis
statistics, the fluid entropy theory, and the minimum enstrophy state.
From some of those initial states,
 we obtain the quasi-stationary states which are close to
the minimum enstrophy state,
but we also find that the quasi-stationary states depend upon initial states, %
even if the initial states have the same energy and
angular momentum, which means the ergodicity does not hold.
}

\begin{document}

\maketitle

%%%%%%%%%%%%%%%%%%%%%%%%%%%%%%%%%%%%%%%%%%%%%%%%%%%%%%%%%
\section{Introduction}
A system with long-range interaction behaves quite differently from that
with short-range interaction because the total energy is not
proportional to the system size and the thermodynamic limit cannot be
defined in an ordinary sense.  In such a case, even a small subsystem of
a large system may not obey the Boltzmann statistics because the
subsystem interacts strongly with the rest of the system and its energy
depends on the size of the whole system.

An example of such systems is pure electron plasma,
 where the electrons interact via
Coulomb's law.  By applying a strong magnetic field to the system, one
can confine the electrons in a container for quite a long time, and
examine quasi-stationary states of the system.

The system can be described by the two-dimensional (2-d) Euler equation with
vorticity of the same sign if the finite Larmor radii of electrons are ignored;
thus, its dynamics is analogous to that of 2-d incompressible
inviscid fluid.
In the 2-d fluid systems, the dynamics of vortices are known to be
important
in understanding the system behavior, as is seen in large-scale
geophysical flows of high Reynolds number.

A system described by the 2-d Euler equation usually develops
a fine vorticity structure by complex 
stretching and folding of the vorticity
field,
which results in finer and finer structures as the system evolves,
but this does not usually lead to a fractal structure in stationary
states; the fine structure eventually becomes too small to observe for
any realistic observation with finite resolution; thus, quasi-stationary
states observed in experiments and simulations usually have a rather
smooth global structure,
which may be predicted by some statistical methods.

Statistical theories of the 2-d perfect fluid originate from
Onsager's work on the point vortex model.
\cite{bstatistical_hydrodynamics}
Since this model is described by a Hamiltonian, 
he has introduced the ``temperature'' in a microcanonical ensemble
by taking the derivative of the logarithm of the number of states
with respect to the total energy.
In particular, in the case of ``negative temperature'',
this theory predicts that small vortices merge together and
 eventually self-organize
into a large single cluster.

In some simulations,
 the connection between the point vortex system and
 Onsager's theory has been
investigated, and the result shows that,
 for the three-particle case,
the initial state in the negative temperature region
leads to the chaotic motion of particles.
\cite{bnpoint_vortex}
A related result that the sign of
the temperature affects the dynamics of the system has been also obtained
for the
system which contains particles with both signs of the vorticity.
\cite{bdynamics_two_sign_point_vortices}

Since Onsager, many theories have been proposed so far.
\cite{btwodim_turbulence_physicist}
Some of them are based on maximization of the entropy,
\cite{
      bnegative_temperature_states,
	  bstatistics_line_vortices,
	  bstatmech_eulereq_jupiter,
      brelaxation_towards_statistical_equilibrium}
while others are
based on minimization of the enstrophy.
\cite{bminimum_enstrophy_vortices}
Most of the theories
 predict the formation of a large single cluster as the equilibrium
state.
However, experimental tests of the theories are difficult 
with an actual 2-d fluid, and the 2-d electron plasma system
can be used to test such theories.

In 1994, an experiment was performed by Huang and Driscoll
using 2-d electron plasma
\cite{btwo_dim_turbulence_exp}.
The results suggests that the quasi-stationary
 state of the 2-d electron plasma
system is described by the minimum enstrophy theory
and cannot be explained by any of the maximum entropy theories.
%which is one of the theories that have been proposed to explain
%the stationary state of 2-d turbulence.
%  In this experiment, density fields
%and statistical observables 
%such as the enstrophy and the entropy are
%defined by coarse graining of the distribution of electron particles.
This is particularly interesting because the minimum enstrophy state
has been shown to be the maximum Tsallis entropy state with
$q = 1/2$ in the Tsallis statistics,
which has been
developed to describe nonextensive systems.
\cite{btwo_dim_turbulence_tsallis}

Numerical simulations on the 2-d electron plasma system
have been performed.
The minimum enstrophy state is also observed in simulation study
using the same initial states as those in the experiment.
\cite{btwodim_euler_unique_final}
However, the initial-state dependence of such a stationary state has not been
tested systematically.
%However, the maximum entropy state is also observed in a simulation of 
%some situation.
The validity of the minimum enstrophy theory has been tested by some authors,
and it was claimed that
the experimentally obtained density profile appears to be the minimum
enstrophy state only because the mixing in the core and peripheral regions
of the vortex cluster has not been completed yet
\cite{bmax_entropy_vs_min_enstrophy}.

Another kind of quasi-stationary state, called the vortex crystal state,
 has been found in experiments.
\cite{btwo_dim_vortex_crystal, bdynamics_statistics_vortex_crystals}
The vortex crystal is 
analogous to the 
coherent vortex structure in 2-d turbulent flow,
\cite{bevolution_of_vortex_statistics}
but the vortex crystal state has not been
described by any known statistical theories;
it is described only phenomenologically.
\cite{bregional_maximum_entropy}
%bsimulation_diocotron_mdrape2

%Numerical simulations on the 2-d electron plasma system
%have been also performed.
In numerical simulations, the existence of the vortex crystal state has been
confirmed and the relaxation process to it has been studied.
It has been shown that the vortex crystal state is 
sensitive to the small difference in initial states, and
is stabilized by background vortices.
 \cite{bvortex_crystals_simulation}
The vortex crystal state is realized not only with a large number of
particles
but also with only several hundred particles.
In the simulation, 
the interaction between vortex clumps and background vortices is
also found to be important; there are exchanges of energy and
particles between them even after the formation of the vortex crystal state.
\cite{bnpoint_vortex}
The effects of background vortices in the vortex crystal state,
 such as the dynamics of
clumps and the formation of a regular structure,
have also been intensively studied in experiments.
\cite{bformation_vortex_crystal_background,
 bdynamics_electron_plasma_background,
 bmerger_binary_background}

In this study, we investigate the quasi-stationary states of the
two-dimensional pure electron plasma under a strong magnetic field
by numerical simulation.
With our numerical results,
we examine several statistical theories for the stationary state.

The paper is organized as follows:
In \S\ref{s_model_system_and_its_behavior}, we introduce
the 2-d electron plasma system under a magnetic field and define
a model.
In \S\ref{s_statistical_theories_for_final_states}, 
we briefly review some statistical theories.
The simulation method and results are described 
in comparison with the statistical theories in \S\ref{s_simulation},
and the summary and
 discussion are given in \S\ref{s_conclusions_and_discussion}.

%%%%%%%%%%%%%%%%%%%%%%%%%%%%%%%%%%%%%%%%%%%%%%%%%%%%%%%%%
\section{Model System and its Behavior}
\label{s_model_system_and_its_behavior}

\begin{figure}[b]
 \begin{center}
   \includegraphics[trim=20 0 0 0,width=0.30\columnwidth]{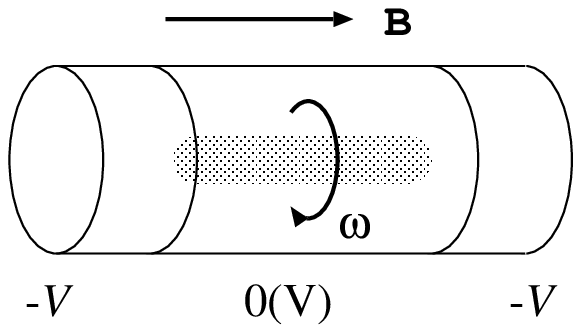}
  \caption{
   Schematic diagram of Malmberg trap of pure electron plasma.}
  \label{f_malmberg}
 \end{center}
\end{figure}

The system we consider is a pure electron plasma system in a
cylindrical container with a uniform external magnetic field along
its axis; in a certain experimental situation,
\cite{
 bdynamics_statistics_vortex_crystals,
 btwo_dim_turbulence_exp,
 btwo_dim_vortex_crystal}
the system behaves like a 2-d one in the plane
perpendicular to the axis; we investigate this situation in the
following.

\subsection{Equations of motion}
\label{ss_equations_of_motion}
We assume that the magnetic field is strong enough for the radii of
cyclotron motion to be ignored; 
then the velocity $\mib{v}_{i}$ of the $i$ - th electron at $\mib{r}_{i}$
within the plane is given by the drift velocity of the guiding center
%$\mib{v}={\mib{E}\times\mib{B}\over B^2} $
\begin{equation}
 \label{m_drift}
 \mib{v}_{i} = \frac{\mib{E}(\mib{r}_{i})\times\mib{B}}{B^2} 
\end{equation}
under the existence of the electric field $\mib{E}(\mib{r})$ perpendicular to
the uniform magnetic field $\mib{B}$.  In the present situation, $\mib{E}$ is
generated by the electrons themselves,
as
\begin{equation}
 \label{m_pointvortex_ef}
 \mib{E} = -\mathrm{grad} \phi,
 \quad
 \nabla^{2}\phi
  = \frac{e}{\epsilon_{0}}\sum_{i}\delta(\mib{r} - \mib{r}_{i}(t)),
\end{equation}
where $\mathrm{grad}$ and $\nabla^{2}$ are the 2-d gradient and
Laplacian,
$\phi$ is the 2-d electric potential with the boundary condition
$\phi = 0$ at the container wall,
$e$ is the charge of an electron, and
$\epsilon_{0}$ is the dielectric constant of vacuum.
Thus, the electrons do not repel
each other and the plasma can be confined in a container (Malmberg
trap, Fig. \ref{f_malmberg}).
The system described by eqs. (\ref{m_drift}) and (\ref{m_pointvortex_ef})
is called the point vortex system, because the interaction between
the electrons is analogous to that between vortices.

If the system contains macroscopic numbers of electrons
and can be described by a smooth density field $n(x,y)$,
then the density field follows the following partial differential equations:
% (\ref{m_euler})$-$(\ref{m_poisson})
%
\begin{equation}
 \label{m_euler}
 \frac{Dn}{Dt} \equiv
 \frac{\partial n}{\partial t} + \mib{v}\cdot \nabla n = 0,
% \mib{v} = \hat{\mib{z}}\times\nabla\phi,
\end{equation}
\begin{equation}
 \label{m_drift_normalized}
 \mib{v} = \hat{\mib{z}}\times\nabla\phi = \left(
  -\frac{\partial \phi}{\partial y}, \frac{\partial \phi}{\partial x}
 \right),
\end{equation}
\begin{equation}
 \label{m_poisson}
\nabla^{2}\phi = n ,
\end{equation}
where $\hat{z}$ denotes the unit vector perpendicular to
the plane and $\nabla$ is the 2-d nabla.  
%*** pattern 1 ***
We have normalized the potential and density as
$(1/B)\phi \mapsto \phi$ and $(e/\epsilon_{0}B)n \mapsto n$.
%where 
%$e$ is the charge of an electron and
%$\epsilon_{0}$ is the dielectric constant of vacuum.
%*** pattern 2 ***
%We have normalized the scale of space and time 
%so that the parameters become unity, as
%$x / L \mapsto x$ and $t / (\epsilon_{0} B L^{2}/ e N) \mapsto t$
%where $L$ is the typical length scale of the system,
%$\epsilon_{0}$ is the dielectric constant of vacuum,
%$e$ is the charge of an electron and
%$N$ is the number of electrons.
%Then the potential $\phi / (e N / \epsilon_{0}) \mapsto \phi$
%and the density $n / (N / L^{2}) \mapsto n$ can be normalized as
%non-dimensional variables.
%the scale of space and time so that 
%the electron charge constant $e$ and the strength of the magnetic field
%$B$ become unity.

It can be shown that the density field $n$ is equal to 
the vorticity $\omega(\mib{r}) \equiv (\nabla \times \mib{v})_{z}$
of the 2-d
velocity field $\mib{v}$ and the velocity field $\mib{v}$ is
solenoidal ($\nabla \cdot \mib{v} = 0$);
therefore, the set of equations (\ref{m_euler})$-$(\ref{m_poisson})
 is the same with that of
the Euler equation for the 2-d incompressible inviscid fluid
with a free-slip (no-stress) boundary condition, but the
vorticity takes only a positive value in the present system.

\begin{figure}[b]
 \begin{center}
  \includegraphics[trim=20 0 0 0,width=0.6\columnwidth]{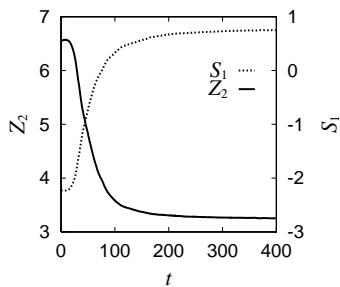}
  \caption{
   Time evolution of two Casimirs in simulation:
   the dotted line and dashed line represent the enstrophy
    $Z_{2}$ and entropy $S_{1}$, respectively.
   The initial distribution is ring shaped with $R_{1} / R_{0} = 0.6$.
   See Fig. \ref{f_phase} for the meaning of the parameter.
   These observables are calculated on the basis of a
   coarse-grained density field
   using a finite-sized mesh with a resolution of $256 \times 256$.
   }
  \label{f_long_fragile}
 \end{center}
\end{figure}

Note that the density field 
\begin{equation}
 n(\mib{r}) = \sum_{i}\delta(\mib{r} - \mib{r}_{i}(t))
\end{equation}
with $\mib{r}_{i}(t)$ being a solution of 
eqs. (\ref{m_drift}) and (\ref{m_pointvortex_ef}) gives a singular solution
for eqs. (\ref{m_euler})$-$(\ref{m_poisson}).

Although the dynamics given by
eqs. (\ref{m_drift}) and (\ref{m_pointvortex_ef})
is not Newtonian, it can be expressed
in the form of canonical equations,
\begin{equation}
 \frac{dx_{i}}{dt} = \frac{\partial H}{\partial y_{i}},
\quad
 \frac{dy_{i}}{dt} = -\frac{\partial H}{\partial x_{i}},
\end{equation}
with the Hamiltonian $H$,
%\begin{equation}
%\label{m_hamiltonian}
%\begin{split}
%  H \equiv
% &-\frac{1}{4\pi}\sum_{i}^{N}\sum_{j\neq i}^{N}
%  \Gamma_{i}\Gamma_{j}\ln |\mib{z}_{i} - \mib{z}_{j}| \\
% &+\frac{1}{4\pi}\sum_{i}^{N}\sum_{j}^{N}
%  \Gamma_{i}\Gamma_{j}\ln |R_{w}^{2} - z_{i}z_{j}^{*}| \\
% &-\frac{1}{4\pi}\sum_{i}^{N}\sum_{j}^{N}
%  \Gamma_{i}\Gamma_{j}\ln R_{w}
% \\
% = &- \frac{1}{2}\sum_{i}^{N}\Gamma_{i}\phi_{i}(\mib{r}_{i})
%\end{split}
%\end{equation} 
\begin{equation}
\label{m_hamiltonian}
\begin{split}
  H
 &= - \frac{1}{2}\sum_{i}^{N}\sum_{j \neq i}^{N}
      G(\mib{r}_{i},\mib{r}_{j})
    - \frac{1}{2}\sum_{i}^{N}
      G_{\mathrm{m}}(\mib{r}_{i},\mib{r}_{i}) \\
 &= - \frac{1}{2}\sum_{i}^{N}\phi_{i}(\mib{r}_{i}),
\end{split}
\end{equation} 
where $G(\mib{r}_{i},\mib{r}_{j})$ is the 2-d Green function
for the electric potential,
$G_{\mathrm{m}}(\mib{r}, \mib{r}')$ is the electric potential
at $\mib{r}$ brought about by the mirror charge 
induced by the charge at $\mib{r}'$ to satisfy the 
boundary condition,
% of a particle at $\mib{r}'$
% invoked by the conductor boundary,
 and $\phi_{i}(\mib{r})$ is the electric
potential due to all the electrons except for the $i$th one.
The Green function satisfies
% the following equation
\begin{equation}
 \nabla^{2}G(\mib{r},\mib{r}') = \delta(\mib{r} - \mib{r}'), 
\end{equation}
under an appropriate boundary condition.
In the present case, we consider that the system is in a cylindrical
 container with the radius $R_{w}$; then $\phi_{i}$ is given by
 \cite{bnpoint_vortex}
\begin{equation}
\begin{split}
  \phi_{i}(\mib{r}) \equiv
 &+\frac{1}{2\pi}\sum_{j\neq i}^{N}
  \ln |\mib{z} - \mib{z}_{j}| \\
 &-\frac{1}{2\pi}\sum_{j}^{N}
  \left[
   \ln |z - \frac{R_{w}^{2}}{z_{j}^{*}}| 
   +\ln \frac{|z_{j}|}{R_{w}}
  \right],
\end{split}
\end{equation}
where $z = x + i y$ and 
$z^{*}$ is the complex conjugate of $z$;
the second term corresponds to the potential brought about by
the mirror charges.

\subsection{Constants of dynamics}
\label{ss_constants_of_the_dynamics}
The Hamiltonian $H$ is a constant of the dynamics, and
is expressed in terms of the field variables as
\begin{equation}\label{m_energy}
\begin{split}
  H 
    &= -\frac{1}{2}\int d^{2}\mib{r}\int d^{2}\mib{r}' \,
        n(\mib{r})n(\mib{r}')G(\mib{r}, \mib{r}') \\
    &= -\frac{1}{2}\int d^{2}\mib{r} \, n(\mib{r})\phi(\mib{r}) \\
    &= \int d^{2}\mib{r} \, \frac{1}{2}\mib{v}^{2}(\mib{r}),
\end{split}
\end{equation}
which corresponds to the total energy.

In the case of a system with circular symmetry,
 the total angular momentum $L$ around the
center of the system,
\begin{equation}\label{m_angular_momentum}
\begin{split}
  L  
  &\equiv \sum_{i}^{N}r_{i}^{2} \\
  &= \int d^{2}\mib{r} \, r^{2}n(\mib{r}) \\
  &= \int d^{2}\mib{r} \, (\mib{r} \times \mib{v}(\mib{r}))_{\mathrm{z}},
\end{split} 
\end{equation}
is also a constant of the dynamics.

In addition to the above two quantities,
if the system is described by a smooth density field that satisfies
eqs.  (\ref{m_euler})$-$(\ref{m_poisson}),
 it can be shown that the
integral $Z$ of any function $f$ of $n$,
\begin{equation}\label{m_casimirs}
  Z \equiv \int d^{2}\mib{r} \, f(n(\mib{r})),
\end{equation}
is a constant of the dynamics; this is called a Casimir constant.
\cite{btwo_dim_turbulence_tsallis, bnonlinear_stability_fluid}

Out of the Casimirs, the enstrophy $Z_{2}$,
\begin{equation}
 \label{m_enstrophy}
  Z_{2} \equiv \frac{1}{2}\int d^{2}\mib{r} \, n^{2}(\mib{r}),
\end{equation}
  and the one-body entropy $S_{1}$,
\begin{equation}
 \label{m_point_vortex_entropy}
  S_{1} = -\int d^{2}\mib{r} \, n(\mib{r}) \log n(\mib{r}),
\end{equation}
are of particular interest.

  Although
any Casimir is a conserved quantity for the dynamics given by
eq. (\ref{m_euler}), it is often not conserved in simulations and
experiments\cite{btwo_dim_turbulence_exp}
  (Fig. \ref{f_long_fragile});
%%, as shown in Fig. \ref{f_long_fragile};
 this is because the field quantities are defined only at a
finite resolution
by averaging over a mesh with a finite size.
 \cite{bvortex_crystals_simulation}

When the density function can be measured
in a fine-enough resolution compared with the density
structure, the value of a Casimir should not
depend on the size of the mesh, but it
depends on the resolution when there exists a density
structure smaller than the mesh size. The present system
often develops very quickly a fine structure in density by
stretching and folding. The smallest
length scale of the structure appears to decrease exponentially
in time, in which case the value of the coarse-grained Casimir
is not conserved as long as the mesh size is finite.

\begin{figure}[b]
 \begin{center}
  \includegraphics[trim=20 0 0 0,width=0.6\columnwidth]{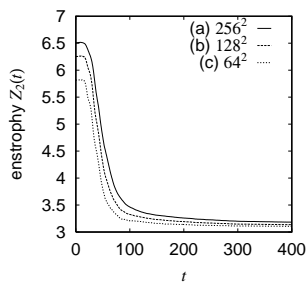}
  \caption{
   Time evolution of enstrophy  $Z_{2}$ of single simulation run
   with several resolutions of coarse-grained mesh:
   The macroscopic initial distribution is the same as
   in Fig. \ref{f_long_fragile}.
   The resolution of the VIC cell necessary to solve Poisson's equation
   is $512\times 512$ (see \S\ref{ss_method} for the meanings of parameters).
   The resolution of density observation is
   $256\times 256, 128\times 128$ and $64\times 64$
   for (a)$-$(c), respectively.
   }
  \label{f_enstrophys_coarse}
 \end{center}
\end{figure}

Note that the density field in the quasi-stationary state,
if coarse grained over these evolving small length scales,
usually has a smooth structure in the scales comparable to the system
size; thus, the value of enstrophy or any Casimir should be well defined
for the coarse-grained field.
This is because the stretching and folding do not lead to a
fractal structure in simulations and experiments;
all the structures except some large structures comparable to the system
size are transfered to finer and finer scales as they evolve; thus,
we find that the scales of these large structures are well separated from
the fine scale in the quasi-stationary state.
Since the fine structure
eventually becomes invisible for any measurement with a finite
resolution, the coarse-grained density field has a fairly simple
 structure with the size 
comparable to the system size, which is independent of
the resolution of measurement as long as the mesh size is finite and
sufficiently smaller than the system size.

This is shown in Fig. \ref{f_enstrophys_coarse}, where
the time developments of enstrophy are plotted using three
different mesh sizes for a single simulation.
The initial and transient values are quite different from each other,
but the values for the quasi-stationary state become closer.
Note that the rate of enstrophy change in the transient is 
almost independent of the mesh size of density coarse graining.

\subsection{Stability and evolution of states}
\label{ss_stability_and_evolution_of_states}
It can be shown that the rotationally symmetric state with a decreasing
density $n$ as a function of $r$ is not linearly unstable
and is numerically stable. On the other hand, the ring state,
where the electrons are distributed in a ring-shaped region,
is linearly unstable, and the instability is
called diocotron instability.
\cite{btheory_nonneutral_plasma}   

\begin{figure}[b]
 \begin{center}
  \raisebox{0.165\columnwidth}{(a)}
  \includegraphics[width=0.18\columnwidth]{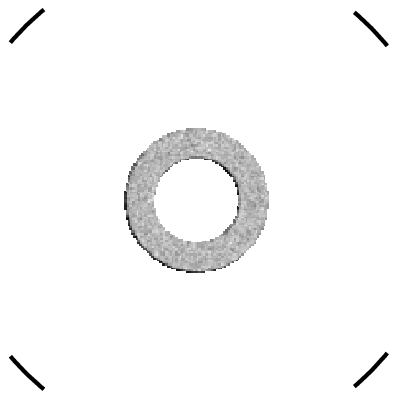}
  \includegraphics[width=0.18\columnwidth]{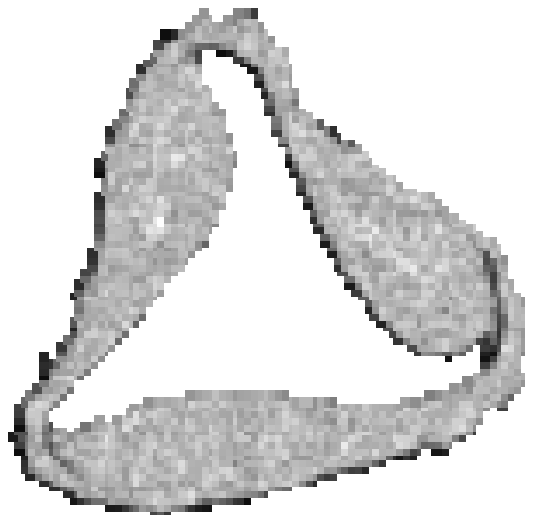}
  \includegraphics[width=0.18\columnwidth]{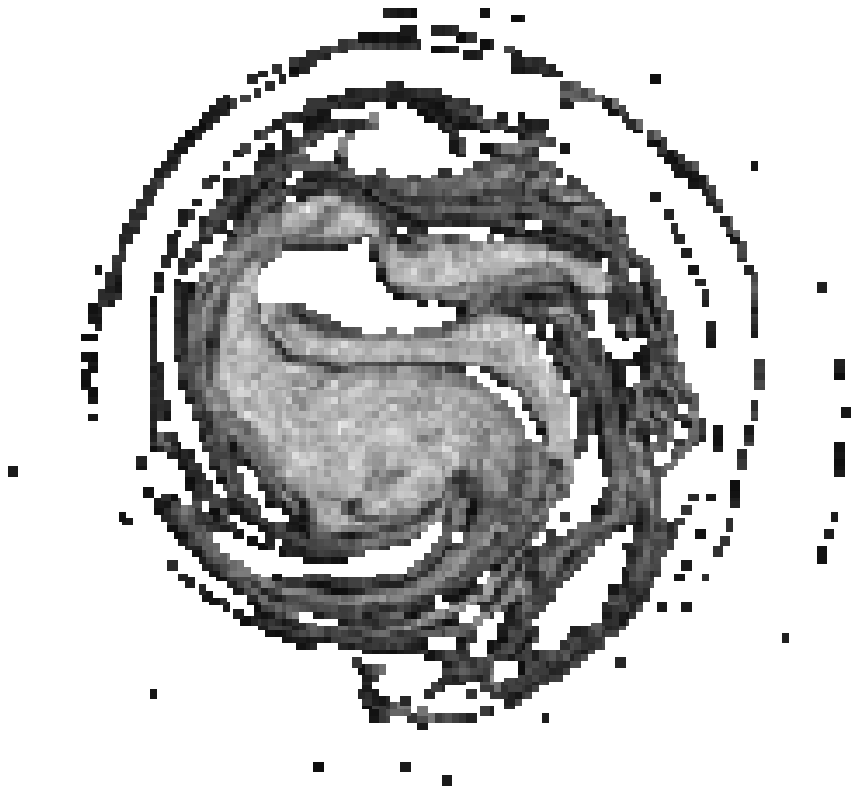}
  \includegraphics[width=0.18\columnwidth]{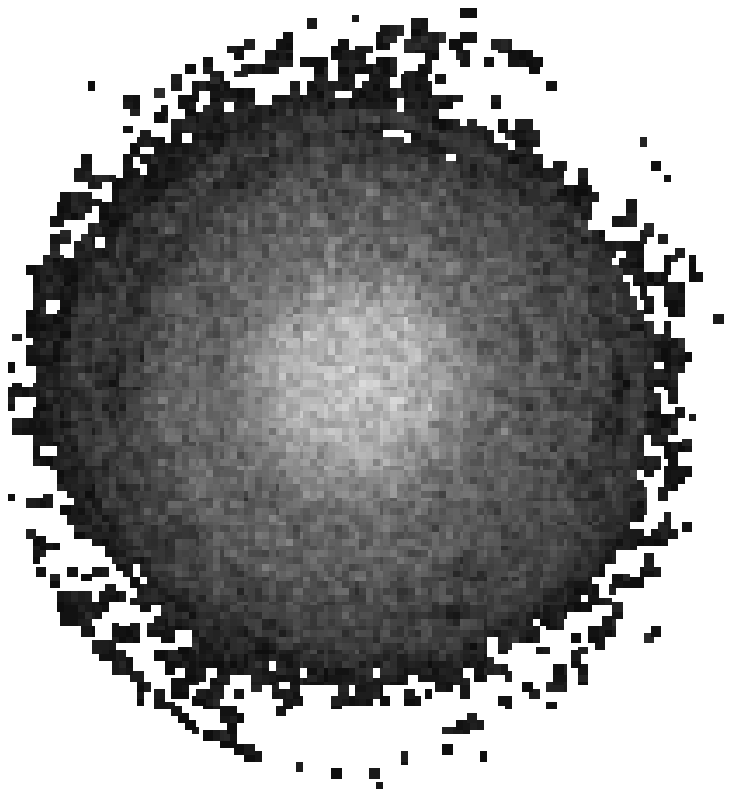}
  \includegraphics[width=0.055\columnwidth]{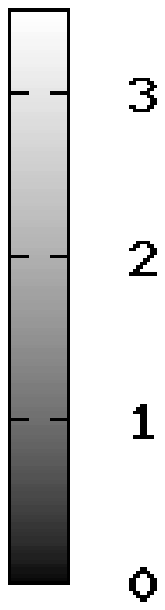}
  \\
  \begin{tabular}[t]{cccc}
   {\small \ $t = 0$\ }
   &
   {\small \ $t = 20$\ }
   &
   {\small \ $t = 60$\ }
   &
   {\small \ $t = 400$\ }
   \\
  \end{tabular}
  \raisebox{0.165\columnwidth}{(b)}
  \includegraphics[width=0.18\columnwidth]{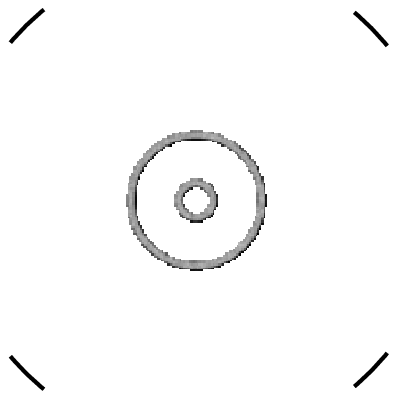}
  \includegraphics[width=0.18\columnwidth]{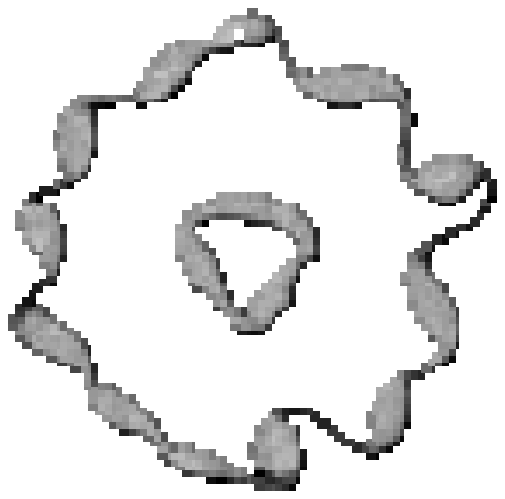}
  \includegraphics[width=0.18\columnwidth]{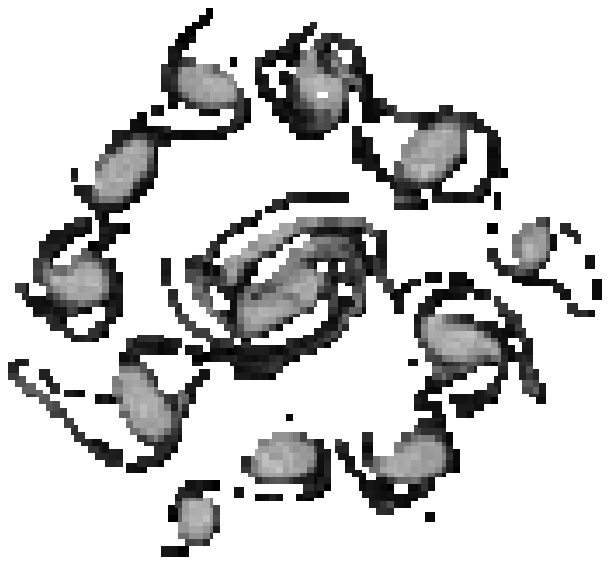}
  \includegraphics[width=0.18\columnwidth]{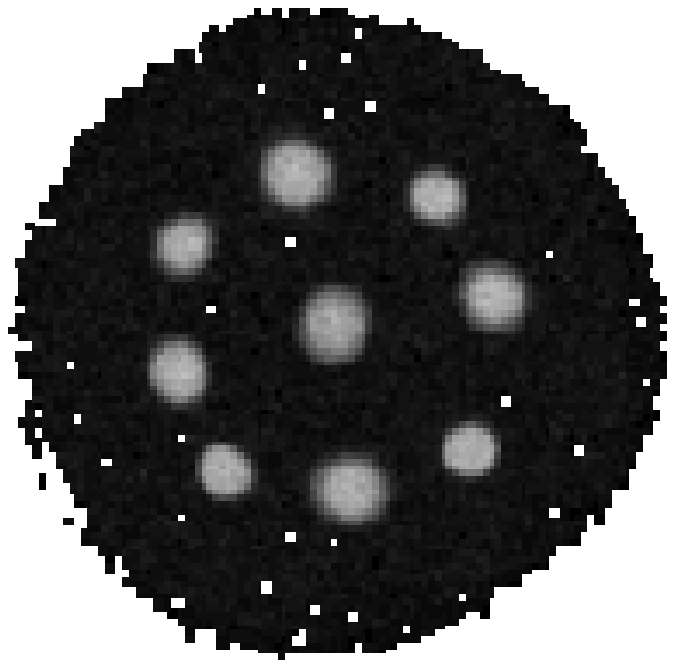}
  \includegraphics[width=0.06\columnwidth]{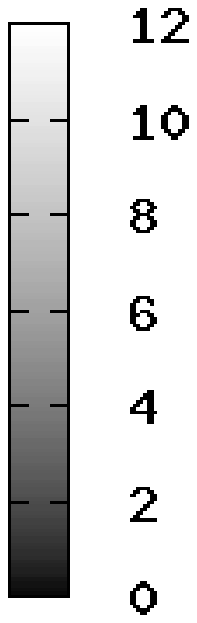}
  \begin{tabular}[t]{cccc}
   {\small \ $t = 0$\ }
   &
   {\small \ \ $t = 4$\ }
   &
   {\small \ \ $t = 10$\ }
   &
   {\small \ $t = 400$\ }
   \\
  \end{tabular}
  \caption{
   Time evolutions of electron density distribution
   from single-ring initial state (a),
   and  double-ring initial state (b).
   Initial distributions are shown in
   the left-most figures and quasi-stationary
   states are shown in the right-most figures.
   The conductor wall boundary is 
   expressed by the black curve in the left-most figure.
   }
  \label{f_animfllsys06hs}
 \end{center} 
\end{figure}

Figure \ref{f_animfllsys06hs} shows two examples of time
sequences which start from unstable initial states and lead to two types
of quasi-stationary states; the sequence in Fig. \ref{f_animfllsys06hs}(a)
 starts from a single-ring
configuration, which destabilizes owing to mode three, and eventually falls
into a singly peaked stable density distribution.  In the case of
Fig. \ref{f_animfllsys06hs}(b),
 a double-ring initial configuration results in instability owing to
a higher mode to break into many vortices, which merge occasionally
while they undergo a collective chaotic motion, and eventually several
surviving vortices form a regular structure, which does not undergo
further change during our simulation time; this state is
called a vortex crystal state.
\cite{btwo_dim_vortex_crystal}

These quasi-stationary states are in general not the equilibrium states.
In our preliminary simulations, it is observed that if there are only
a small number of particles (for example, several hundreds), 
the system evolves very slowly in time after the quasi-stationary
states are attained.
 However, for a large number of particles as in experiments and
current simulations, time evolution after entering 
the quasi-stationary state is
hardly observed.

%%%%%%%%%%%%%%%%%%%%%%%%%%%%%%%%%%%%%%%%%%%%%%%%%%%%%%%%%
\section{Statistical Theories for Quasi-Stationary States}
\label{s_statistical_theories_for_final_states}

As explained in \S\ref{ss_constants_of_the_dynamics}, 
we usually observe the system through the coarse-grained mesh
because we are interested in the macroscopic structures of the system.

After the complex stretching and folding of the density structure,
only a fine structure and a smooth structure are left in
the ``true'' density distribution at the late stage of the evolution.
Thus, the coarse-grained density field
reaches the quasi-stationary states,
which only has the smooth structure of a simple
density field.

Although the dynamics of the present system is highly nonlinear and
unpredictable, quasi-stationary states may be predicted by a simple principle;
actually, there have been several statistical theories
developed to predict the vortex cluster as a quasi-stationary state using
variational methods.
\cite{btwodim_turbulence_physicist}
 We briefly review some of them
in this section.

%% ------------------------------------------------------
\subsection{Maximum one-body entropy state}
Although the one-body entropy $S_1$ (eq. (\ref{m_point_vortex_entropy}))
is a Casimir constant, 
in actual simulations,
the coarse-grained one-body entropy
shows a rapid initial increase and reaches a steady value
(Fig. \ref{f_long_fragile});
thus, the coarse-grained 
density distribution $n$ of the steady state might be given by
maximizing $S_1$, keeping the total number of electrons $N$, the total
energy $H$, and the total angular momentum $L$ constant.
Using the variational method,
\begin{equation}
 \label{m_maxent_variational}
 \delta (S_{1} - aN - bH - cL) = 0,
\end{equation}
where $a, b$ and $c$ are Lagrange multipliers and determined from
the constraints.\cite{btwo_dim_turbulence_exp}
If the energy $H$
of eq. (\ref{m_energy}) is evaluated ignoring electron correlation,
%Substituting equations
% (\ref{m_point_vortex_entropy}), 
% (\ref{m_energy}) and
% (\ref{m_angular_momentum}) into eq.(\ref{m_maxent_variational}),
%\begin{equation}
%\begin{split}
%\delta [
%  &-\int d^{2}\mib{r} \, n(\mib{r}) \log n(\mib{r}) \\
%  &-a\int d^{2}\mib{r} \, n(\mib{r}) \\
%  &+b\frac{1}{2}\int d^{2}\mib{r}\int d^{2}\mib{r}' \,
%        n(\mib{r})n(\mib{r}')G(\mib{r}, \mib{r}') \\
%  &-c\int d^{2}\mib{r} \, r^{2}n(\mib{r})]
% = 0 .
%\end{split}
%\end{equation}
then we obtain
\begin{equation}
 n(\mib{r})
  = \exp(-1 - a + b \phi(\mib{r}) - c r^{2}).
\end{equation}
From  eq. (\ref{m_poisson}), $\phi$ satisfies
\begin{equation}
 \nabla^{2}\phi(\mib{r})
  = \exp(-1 - a + b \phi(\mib{r}) - c r^{2}),
\end{equation}
which can be solved numerically .

\subsection{Maximum fluid entropy state}
The fluid entropy
\cite{bstatmech_euler_equations,
      bstatmech_eulereq_jupiter,
      brelaxation_towards_statistical_equilibrium}
 is the entropy defined in terms of the probability
distribution $p(\sigma_{j};\mib{r}_i)$ of the
discretized vorticity
$\sigma_{j}$ in the coarse-grained cell at $\mib{r}_i$ as
\begin{equation}\label{m_fluid_entropy}
S_{\rm fluid}\equiv -\sum_{\mib{r}_i} \sum_{\sigma_{j}} 
           p(\sigma_{j};\mib{r}_i)\ln p(\sigma_{j};\mib{r}_i),
\end{equation}
where $\sigma_{j}$'s are the strengths of microscopic vortex elements.
This fluid entropy has been introduced on the basis of the idea of
counting the number of configurations of vortices over
discretized space.
\cite{bstatmech_euler_equations}

The probability distribution $p(\sigma_{j};\mib{r}_i)$ satisfies
the normalization
\begin{equation}
 \label{m_fluid_probability_normalization}
 \sum_{\sigma_{j}}  p(\sigma_{j};\mib{r}_{i})
  = 1
\end{equation}
at each $\mib{r}_{i}$, and is related to the total numbers
of vortices $g(\sigma_{j})$ of the strength  $\sigma_{j}$ by
\begin{equation}
 \label{m_fluid_gsigma}
 g(\sigma_{j})
  = \sum_{\mib{r}_{i}} p(\sigma_{j};\mib{r}_{i});
\end{equation} 
it is assumed that $g(\sigma_{j})$ does not change because 
the strength of each vortex does not change.

The coarse-grained 
density (vorticity) $n(\mib{r}_i)$ at the cell $\mib{r}_i$ is given by
\begin{equation}
 \label{m_coarsegrain_n}
 n(\mib{r}_i) = \sum_{\sigma_{j}} p(\sigma_{j};\mib{r}_i) \sigma_{j} .
\end{equation}

The maximum fluid entropy state is given by maximizing
eq. (\ref{m_fluid_entropy}) over a variation of $p(\sigma_{j};\mib{r}_i)$
under the constraints of eqs. (\ref{m_fluid_probability_normalization})
and (\ref{m_fluid_gsigma}) in addition to $H$ and $L$.
%Under the variation, the total number of vortices with each strength
%\begin{equation}
% g(\sigma_{j})
%  = \sum_{\mib{r}_{i}} p(\sigma_{j};\mib{r}_{i})
%\end{equation} 
%and the normalization factor of probability at each position
%\begin{equation}
% \nu(\mib{r}_{i}) = \sum_{\sigma_{j}}  p(\sigma_{j};\mib{r}_{i})
%  = 1
%\end{equation}
%are
%constant in addition of $H$, and $L$ (total number of particles $N$ is
%automatically conserved by these constraints). Therefore,
%the variational equation becomes
%\begin{equation}
% \delta\left(S - bH - cL
%             - \sum_{\sigma_{j}} \gamma(\sigma_{j})g(\sigma_{j})
%             - \sum_{\mib{r}_{i}} \kappa(\mib{r}_{i})\nu(\mib{r}_{i})
%             \right) = 0,
%\end{equation}
%where $b, c, \gamma(\sigma_{j})$ and $\kappa(\mib{r}_{i})$ are again
%Lagrange multipliers and determined by the constraints.
This gives
\begin{equation}
  p(\sigma_{j};\mib{r}_{i}) 
 = \frac{\exp[-\gamma(\sigma_{j}) + b\phi(\mib{r}_{i}) - cr_{i}^{2}]}
        {\sum_{\sigma_{k}}
         \exp[-\gamma(\sigma_{k}) + b\phi(\mib{r}_{i}) - cr_{i}^{2}]},
\end{equation} 
where $\gamma(\sigma_{j}), b$ and $c$ are Lagrange multipliers.
Note that the total number of particles $N$ is given by
\begin{equation}
N = \sum_{\sigma_{i} \neq 0} g(\sigma_{i});
\end{equation}
therefore, we need not include $N$ as a constraint.
In the continuum limit in space, combining this with 
eqs. (\ref{m_coarsegrain_n}) and (\ref{m_poisson}),
 the potential $\phi$ can be solved numerically.

In principle, these constants and the values of $\sigma_{j}$'s
should be determined by the initial states of 2-d fluid.
However, there is ambiguity in the values of  $\sigma_{j}$'s since
we only know the coarse-grained information of the initial states,
 $n(\mib{r})$, while different sets of $\sigma_{j}$'s can
 give the same $n(\mib{r})$.
To apply the theory to our simulation of 2-d electron plasma,
we take the values of $\sigma_{j}$'s to be those of the
coarse-grained density of the initial states,
as Huang and Driscoll did in their analysis of the
experiment.\cite{btwo_dim_turbulence_exp}

\subsection{Minimum enstrophy state}
Another Casimir constant, the enstrophy $Z_2$
defined by eq. (\ref{m_enstrophy}),
is known to
cascade towards smaller length scales,
\cite{binertial_twodim, bfree_enstrophy_cascade}
 and be dissipated in the regime where
eq. (\ref{m_euler}) is not exact; numerically, the coarse-grained
enstrophy has been shown to
decrease initially until it reaches a steady value; thus,
coarse-grained density field
 $n$ for the steady states could be given by minimizing $Z_2$ within
given values of $N$, $H$ and $L$ as in the maximum entropy state.
\cite{bminimum_enstrophy_vortices, btwo_dim_turbulence_exp}
%\begin{equation}
% \delta(Z_{2} - a N - b H - c L) = 0
%\end{equation}
This gives
\begin{equation}
 \label{m_minenst_linear}
 n(\mib{r}) = a -b\phi(\mib{r}) + cr^{2} ,
\end{equation}
where $a, b,$ and $c$ are again the Lagrange multipliers and determined
by the constraints.
Operating the 2-d Laplacian
 on both-hand sides of eq. (\ref{m_minenst_linear})
and assuming $n(\mib{r})$ is axisymmetric, we obtain
\begin{equation}
\label{m_minenst_besseleq}
\frac{d^{2}n}{dr^{2}}
 + \frac{1}{r}\frac{dn}{dr}
 + bn - 4c = 0.
\end{equation}
This is the 0th-order Bessel equation; therefore,
the solution is given by the Bessel function.
The problem, however, is that the Bessel function takes values with the
both signs but the density $n(\mib{r})$ should be always positive or
zero. This means that the function that minimizes $Z_{2}$ is outside the
domain of physically allowed functional space. In such a case, the
function that minimizes $Z_{2}$ within the physical domain is the one at
the edge of the physical domain in the functional space, i.e., the
function that satisfies eq. (\ref{m_minenst_besseleq}) for the section of
$r$ where the solution is positive but is zero where it is negative.

%The problem, however, is that the Bessel function takes negative
%values but the density $n(\mib{r})$ should not.
%This means that the solution of the variational
%method (eq.(\ref{m_minenst_linear})) does not give the physically
%accessible minimum; due to the restriction of $n \ge 0$,
%the boundary $n = 0$ is also the candidate of the function $n(\mib{r})$
%which gives the minimum of $Z_{2}$ at some region of $\mib{r}$.
%The physical solution, which minimizes $Z_{2}$ within the physical
%domain, is given by the solution of eq.(\ref{m_minenst_besseleq})
%with $n$ being set 0 where $n < 0$.

%but the density $n(\mib{r})$ cannot take negative values.
%In such a case, the minimal is not the minimum which is described by
%eq.\ref{m_minenst_linear} and the minimal will be at the edge of the
%domain, $n(\mib{r}) = 0$ for some region of $r$.

For $b > 0$, which corresponds with the higher energy case,
we employ the solution
\begin{equation}
 n(r) = \left\{
\begin{array}{ll}
  \alpha [J_{0}(\beta r) - J_{0}(\beta r_{0})], & (r < r_{0}) \\
  0,  & (r_{0} < r) 
\end{array} 
\right.
\end{equation}
where $J_{0}(x)$ is the 0th-order Bessel function, and
 $\alpha, \beta$, and $r_{0}$ are determined by the constraints.
The parameter $r_{0}$ is a cutoff length, which is 
introduced to avoid a negative density region,
\cite{bminimum_enstrophy_vortices, btwo_dim_turbulence_exp}
where it is assumed that the region of $n = 0$ is only in the
outer part of the system, $r_{0} < r$. This type of solution is called
the restricted minimum enstrophy model
\cite{btwo_dim_turbulence_exp}.

For $b < 0$ (the lower energy case), the solution is
\begin{equation}
 n(r) = \left\{
\begin{array}{ll}
  \alpha [I_{0}(\beta r) - I_{0}(\beta r_{0})], & (r < r_{0}) \\
  0,  & (r_{0} < r)
\end{array} 
\right.
\end{equation}
where $I_{0}(x) = J_{0}(ix)$ is the 0th-order modified Bessel function.
This solution gives the pancake shape of $n(\mib{r})$
in the limit of $\beta \rightarrow \infty$
($\alpha < 0$), which corresponds to the minimum energy state.

Actually, it has been shown experimentally that there are some cases
in which
quasi-stationary states are very close to the minimum enstrophy states.
\cite{btwo_dim_turbulence_exp}

It has been pointed out that the minimum enstrophy state is 
the state that maximizes Tsallis entropy
\begin{equation}\label{m_tsallis_entropy}
  S_{q} = \frac{1}{1 - q}\int d^{2}\mib{r} \,
    \left[p^{q}(\mib{r}) - p(\mib{r})\right]
\end{equation}
with $q=1/2$ in the Tsallis statistics formalism.
\cite{btwo_dim_turbulence_tsallis}

%%%%%%%%%%%%%%%%%%%%%%%%%%%%%%%%%%%%%%%%%%%%%%%%%%%%%%%%%
\section{Simulations}
\label{s_simulation}
%% ------------------------------------------------------
\subsection{Method}
\label{ss_method}
We simulate 
eqs. (\ref{m_drift}) and (\ref{m_pointvortex_ef})
 using a variation of the point vortex method
\cite{bvortex_method_flow_simulations};
 for
a given set of electron (or vortex) positions, we calculate the electron
density $n(\mib{r})$ on a grid point by averaging over the cell; then
Poisson's equation is solved to obtain the electrostatic potential
$\phi(\mib{r})$, from which the electric field $\mib{E}(\mib{r})$ is
calculated, and the electron positions are updated by
 eq. (\ref{m_drift});
this is called the vortex-in-cell (VIC) simulation.
\cite{bvortex_crystals_simulation, bplasma_physics_via_computer}
Numerical integration in time is performed
 using the second-order Adams-Bashforth-Adams-Moulton method
and Poisson's equation is solved by an implicit method adopting
multiple grids to accelerate the convergence.
\cite{bnumerical_recipe}

\begin{figure}[b]
 \begin{center}
  \includegraphics[trim=20 0 0 0,width=0.6\columnwidth]{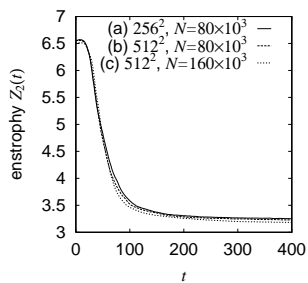}
  \caption{
   Time evolution of enstrophy  $Z_{2}$ for different VIC cell
   resolutions and numbers of particles:
   The macroscopic initial distribution is the same as
   in Fig. \ref{f_long_fragile}.
   For these three simulations, the density field is observed in
   the $256\times 256$ mesh.
   The resolution of the VIC cell necessary to solve Poisson's equation and
   the number of particles are
   (a) $256\times 256$ and $N=80\times 10^{3}$,
   (b) $512\times 512$ and $N=80\times 10^{3}$, and
   (c) $512\times 512$ and $N=160\times 10^{3}$, respectively
   }
  \label{f_enstrophys_meshcomp}
 \end{center}
\end{figure}

The effects of these simulation parameters, such as the resolution of
the VIC
cell necessary to solve Poisson's equation and the number of particles,
are shown in Fig. \ref{f_enstrophys_meshcomp}.
We see almost no difference among these three simulations of
different VIC cell resolutions and numbers of particles.
Thus, for singly peaked quasi-stationary states, simulation results
are almost independent of these parameters.

For initial states which lead to vortex crystal states, however,
the systems are sensitive to the microscopic difference of the 
initial states and may lead to macroscopically different density
distributions of vortex crystal
states because of the chaotic motion in the transient stage.
In this case, the simulation results also depend on those simulation
parameters, such as the VIC cell resolution, the number of particles
and the random number sequence of the initial distribution.
We do not focus on the sensitivity
in the density distribution of vortex crystals
since our aim on this study is to compare singly peaked 
quasi-stationary states of simulations with the statistical theories.

The following simulations are performed on the $256\times 256$
grid with $N=80\times 10^3$ particles.

%% - - - - - - - - - - - - - - - - - - - - - - - - - - - - - - -
\subsection{Results}
\subsubsection{General trends and classification of quasi-stationary states}
\label{sss_general_trends}

\begin{figure}[b]
 \begin{center}
  \raisebox{0.19\columnwidth}{(a)}
  \includegraphics[trim=10 0 0 0,width=0.20\columnwidth]{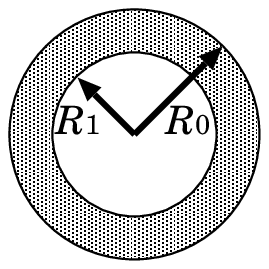}
  $\quad$
  \raisebox{0.19\columnwidth}{(b)}
  \includegraphics[trim=10 0 0 0,width=0.60\columnwidth]{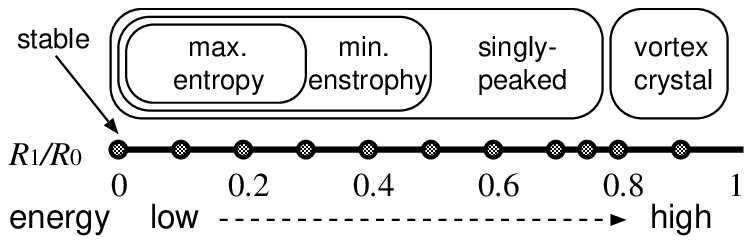}
  \caption{
   (a) Initial ring state.
   (b) Schematic diagram of quasi-stationary states for various ring widths
      $R_{1} / R_{0}$.}
  \label{f_phase}
 \end{center}
\end{figure}

\begin{figure}[b]
 \begin{center}
  \includegraphics[trim=30 0 75 10,width=0.352\columnwidth,clip]{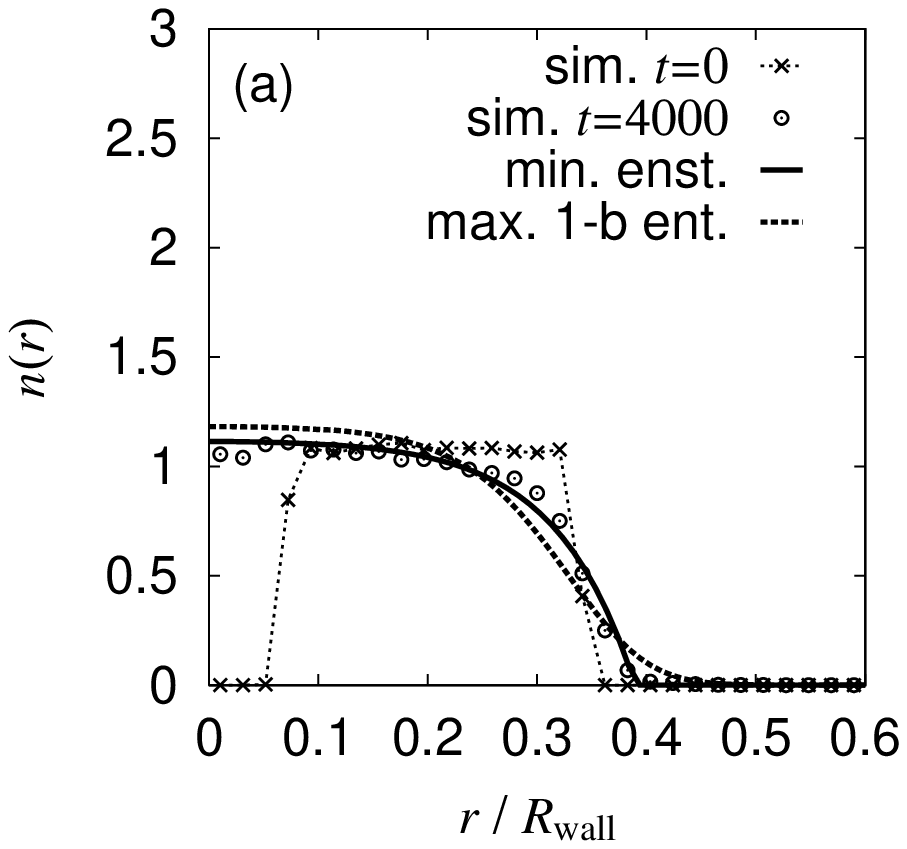}
  \includegraphics[trim=62 0 75 10,width=0.31\columnwidth,clip]{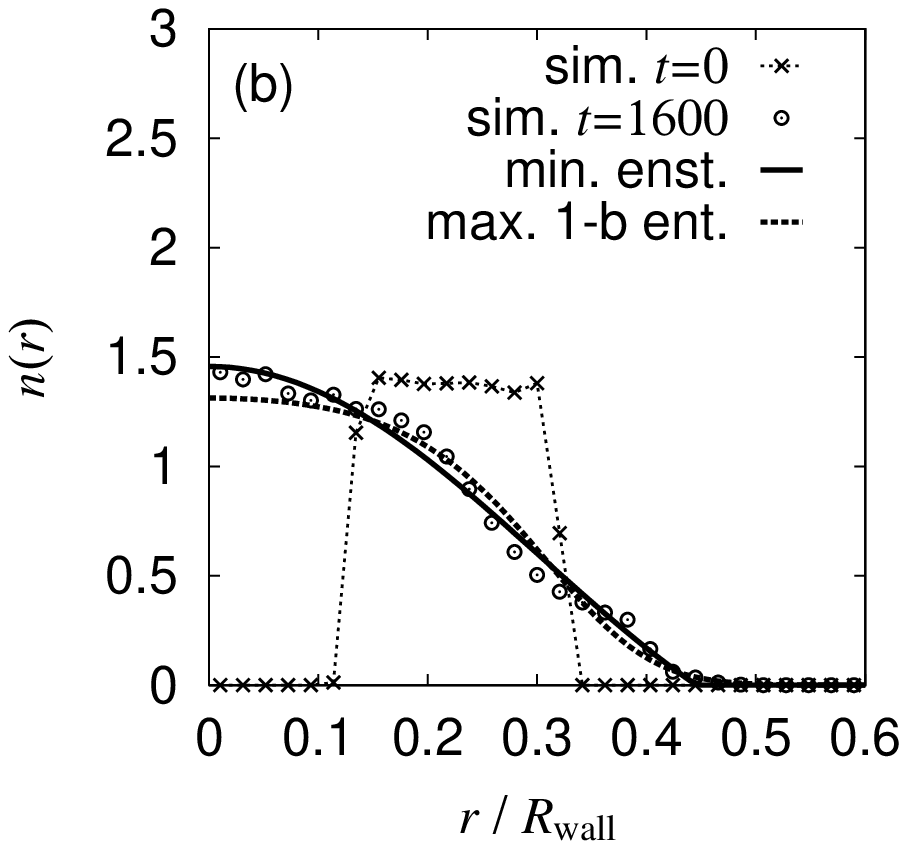}
  \includegraphics[trim=62 0 75 10,width=0.31\columnwidth,clip]{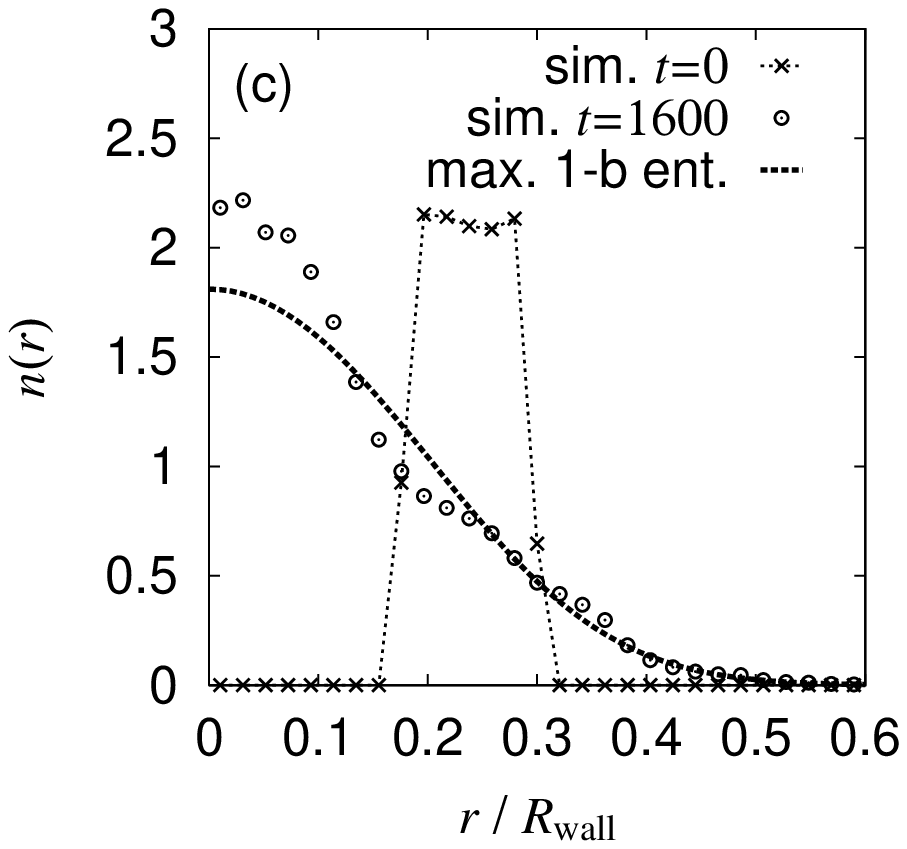}
  \caption{
   Radial density distributions for stationary states from
   initial ring states with
   $R_{1}/R_{0} = 0.2$ (a), $0.4$ (b) and $0.6$ (c). 
   The value of the angular momentum is $L = \pi / 2$ for all three cases, and
   the energies are $E = 2.062$ (a), $2.078$ (b) and $2.113$ (c).
   The minimum enstrophy state and maximum one-body entropy state
   are also shown except for $R_{1}/R_{0} = 0.6$ (c), where
   no physical solutions of the minimum enstrophy state exist.}
  \label{f_ringfinal}
  \end{center}
\end{figure}

We have systematically examined quasi-stationary states in the case where 
the system
starts from the ring states with various ring
widths (Fig. \ref{f_phase}(a));
 the
initial ring state is the state where the electrons are distributed
uniformly in the region $r\in[R_1,R_0]$, and the ratio $R_1/R_0$ is
changed from 0 to 1.  The total number of electrons $N$ and the angular
momentum $L$ are fixed to $N=80\times 10^{3}$ and $L= \pi /2$
 in all cases, but the
energy $H$ is an increasing function of $R_1/R_0$; namely, the energy is
larger for the thinner ring.

The results are summarized in Fig. \ref{f_phase}(b).

In the case of 
the thinner-ring initial state with $0.8\lesssim R_1/R_0 \leq 1$, after
the diocotron instability, the system goes through the chaotic motion of
electron clumps; the clumps merge occasionally and finally result in the
vortex crystal state, which does not change any more.  The number of
clumps in the quasi-stationary
state is very sensitive to the initial configuration
because the clumps merge accidentally during chaotic motion; their
final number depends even upon the microscopic positions of electrons in
the initial states.

For smaller values
of $R_1/R_0 \lesssim 0.8$, the quasi-stationary states are singly
peaked stationary states with circular symmetry; in this regime, the
quasi-stationary state is rather insensitive to the details
of the initial state and
does not depend upon the microscopic
configuration.
  As the ratio $R_1/R_0$
decreases, the density distribution of the quasi-stationary state becomes flatter
around the peak at the center, as can be seen in
Fig. \ref{f_ringfinal},
 the initial ``pancake state''
 with  $R_1/R_0=0$ is stable, and the system does not evolve at all.

For $R_1/R_0\lesssim 0.5$, the minimum enstrophy state seems to give a
reasonably good description of the quasi-stationary state
(Fig. \ref{f_ringfinal}(b)), and for
$R_1/R_0\lesssim 0.3$, the minimum enstrophy state and maximum
entropy state are not so different and both of them are close to
the quasi-stationary state obtained from the simulations.
(Fig. \ref{f_ringfinal}(a))

Note that the minimum
enstrophy state has  
no physical solution with the present angular momentum
for $R_{1}/R_{0}\gtrsim 0.5$.\cite{btwo_dim_turbulence_exp}

Actually, for $R_1/R_0\approx 0$, all the statistical theories give
the states that resemble each other very much and cannot be distinguished
clearly; this is the result of the fact that, for $R_1/R_0=0$, the only
possible state for the given energy and angular momentum is the initial
pancake state; thus, all the statistical theories should give the same
state with this initial state.

The minimum enstrophy state seems to give a better approximate state for
$0.3 \lesssim R_{1}/R_{0} \lesssim 0.5$ than the maximum entropy state.
This is because the simulations give the quasi-stationary state
with a steeper peak
for a larger $R_1/R_0$ and the minimum enstrophy state tends to give the
state with a steeper peak in $n$ than the maximum entropy state
(Fig. \ref{f_ringfinal}(b) and \ref{f_ringfinal}(c)).

On the other hand, it should be noted that there is a discrepancy
between the minimum enstrophy states and the stationary states obtained
by the simulations, if one compares them carefully;
in the minimum enstrophy state, the tail of the density profile is cut
off with a finite slope at a certain point, beyond which the density is
zero, while in the simulations, the density goes to zero smoothly with
the zero slope.

%% - - - - - - - - - - - - - - - - - - - - - - - - - - - - - - -
\subsubsection{Ergodicity}
\begin{figure}[b]
 \begin{center}
  \includegraphics[trim=30 0 75 0,width=0.42\columnwidth]{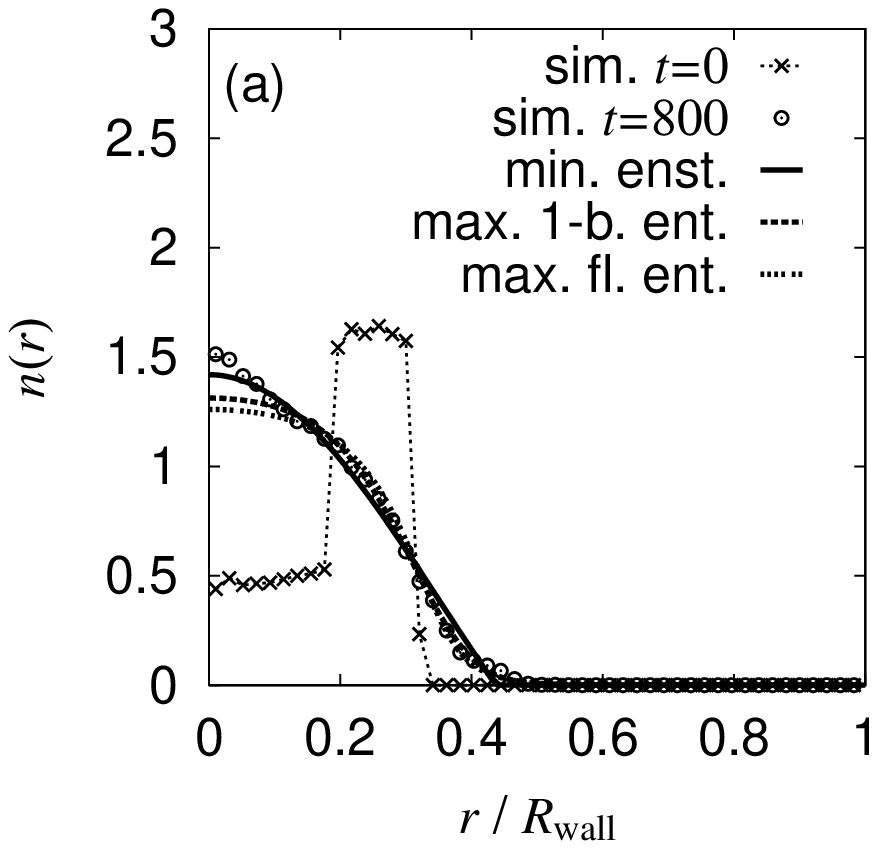}
  \includegraphics[trim=30 0 75 0,width=0.42\columnwidth]{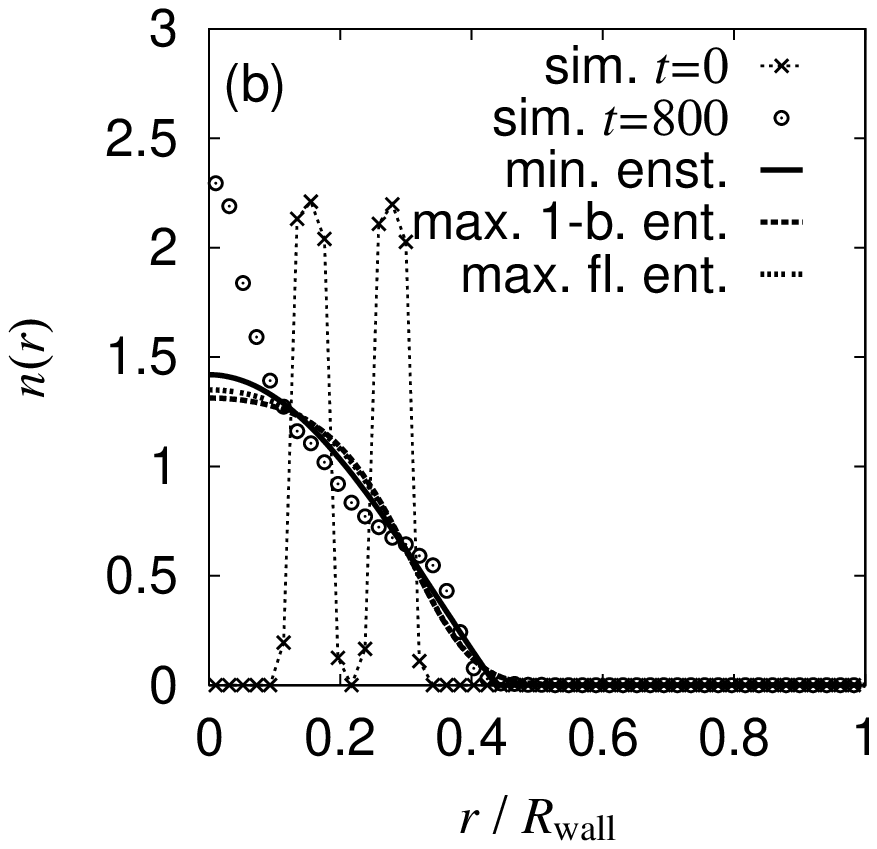}
  \caption{%
   Two different quasi-stationary
   states from two different initial states with the
   same energy $H = 2.078$ and angular momentum $L = \pi / 2$:
   The initial states are the single ring with a finite interior
   density (a) and the double ring (b). The minimum enstrophy state,
   maximum one-body entropy state and maximum fluid entropy state
   are also
   shown; they are the same for both cases except for the maximum
   fluid entropy states; the maximum fluid entropy depends on
   the set of values $\sigma_{j}$, which is different for the two cases:
   $\sigma_{0} = 0$, $\sigma_{1} = 0.49$ and $\sigma_{2} = 1.62$
    for (a), and
   $\sigma_{0} = 0$ and $\sigma_{1} = 2.25$ for (b).
   }
  \label{f_bllsys06as_comp}
 \end{center}
\end{figure}

Although the minimum enstrophy state generally gives 
 a reasonable description for
$R_1/R_0\lesssim 0.5$ in the case of the ring initial state,
the quasi-stationary states are not uniquely determined by the
energy and angular momentum.
The system can travel over only a subsection of the phase space at
given values of energy and momentum; that is, the system is not ergodic.

This can be seen in Fig. \ref{f_bllsys06as_comp}, which shows the
 quasi-stationary  states and the states given by the
statistical theories for the two initial states with the same values of
$N$, $L$
and $H$, namely, the single-ring initial state with a finite density
inside the ring 
(Fig. \ref{f_bllsys06as_comp}(a)),
  and the double-ring
initial state (Fig. \ref{f_bllsys06as_comp}(b)).

When we try to fit these states to the ones from the statistical
theories, we should note that
the maximum one-body entropy states and minimum enstrophy states
are the same for the two cases because they only depend on the values of
$N, L$ and $H$.
The quasi-stationary states obtained by the simulations are
clearly different,
and the one from the single-ring-like state almost coincides with
the minimum enstrophy state,
while
the quasi-stationary state from the double ring
is different from both of the maximum entropy
and minimum enstrophy states mainly because of
the steep peak at the center.

To determine if there are any statistical theories
that fit the quasi-stationary states from the double ring,
we try the maximum fluid entropy state and
maximum Tsallis entropy state, because
these theories contain extra parameters other than
the energy and angular momentum; therefore,
they may be able to distinguish the two initial states.

%Fig.\ref{f_bllsys06as_comp} also shows the density profiles for the
%maximum fluid entropy state.

The maximum fluid entropy state depends on the choice of a set of $\sigma_j$.
In Fig. \ref{f_bllsys06as_comp},
 we use the values determined from the initial density as shown in the
figure caption.
The resulting state, however, does not agree well 
with either of the stationary states as can be seen in  
Fig. \ref{f_bllsys06as_comp}.

\begin{figure}[b]
 \begin{center}
  \includegraphics[trim=30 0 75
  0,width=0.42\columnwidth]{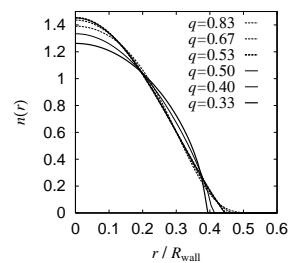}
 \end{center}
 \caption{Maximum Tsallis entropy states for various values of $q$
 with the same energy $H = 2.078$ and angular momentum
 $L = \pi / 2$, which correspond to parameters of
 Fig. \ref{f_bllsys06as_comp}.}
 \label{f_getest_comp-m}
\end{figure}

Figure \ref{f_getest_comp-m} shows the density profiles for
the maximum Tsallis entropy states.
A maximum Tsallis entropy state has a parameter $q$; the state with
$q=1/2$ is the minimum enstrophy state and that with $q = 1$ is the maximum
one-body entropy state.
It can be seen that the state with a smaller value of $q$, such as
$q=1/3$ state, shows a fatter profile and is not closer to the simulation
data in Fig. \ref{f_bllsys06as_comp}(b) 
than the $q=1/2$ state, or the minimum enstrophy state.  Actually,
one can see that the state with $q\approx 0.53$ shows the steepest density
profile, which turns out to be very close to the minimum enstrophy state.
It has been shown \cite{btwodim_turb_pep_nonextensive}
that the peak density decreases monotonically as $q$ increases
in the parameter region $0.5 < q < 1.0$.
Again, it is clear that we cannot fit the state to a
maximum Tsallis entropy state with any choice of the parameter $q$.

%---------------------------------------------------------------------
\subsubsection{Peak density in quasi-stationary state}
\label{sss_peak_density}
There exists a high-density region around the center in the stationary
state from the double ring in Fig. \ref{f_bllsys06as_comp}(b).
This peaky profile cannot be fitted by any of the
statistical theories we tried.
%The major reason that we cannot fit the stationary state
%from the double ring in Fig.\ref{f_bllsys06as_comp} to
%any statistical theories we tried is that there exists a
%high density region around the center.
The value of the peak density in the quasi-stationary
state is  approximately the same as
that in the initial state. This property
can be seen in other cases of
 Figs. \ref{f_ringfinal} and \ref{f_bllsys06as_comp}
and in almost all the simulations we have carried out.

From the time developments of the states, it is
conceivable that, during the relaxation process, 
the initial density is kept in the region of the peak density.
The peak density in quasi-stationary states is never larger than
that of initial states.

A similar feature has been observed in the vortex merger process
 in two-dimensional free decaying turbulence,
where the vortices of both charges exist.
\cite{bevolution_of_vortex_statistics, bwhich_vortex_is_victorious}

\section{Summary and Discussion}
\label{s_conclusions_and_discussion} 

%We have performed the numerical simulations on the two-dimensional Euler
%equation with non-negative vorticity field, in order to study the quasi stationary
We have performed the numerical simulations on the two-dimensional point
vortex model with a unit circulation of the same sign,
 in order to study the quasi-stationary
 states of pure electron plasma under a strong magnetic
field, and have compared the coarse-grained density distribution
of the quasi-stationary state with those of several statistical theories.
By changing the radii ratio $R_1/R_0$ of the initial ring state, we found
(i) for $0.8\lesssim R_1/R_0 \leq 1$,  the quasi-stationary state is 
a vortex crystal state and
(ii) for $0 \lesssim R_1/R_0 \leq 0.8$, the quasi-stationary state is a singly
peaked state.
Further study in the parameter region of the singly peaked state revealed
that
(iii) for $0 \lesssim R_1/R_0 \leq 0.5$, the 
quasi-stationary singly peaked state
is close to the minimum enstrophy state and
(iv) for $0 \lesssim R_1/R_0 \leq 0.3$, the maximum entropy state and
minimum enstrophy state are close to each other,  and the quasi-stationary
state is close to both of them.

These are consistent with the experiment,
\cite{btwo_dim_turbulence_exp}
where the quasi-stationary state
has been found to be very close to the minimum enstrophy state when the
initial state is prepared as the ring state.

Although these findings may provide evidence that the minimum
enstrophy state is a good candidate for the statistical theory that can
describe
the quasi-stationary state for some parameter region,
 we have also found that this system lacks
ergodicity; namely, the system does not necessarily fall into the same
state even though its initial states have the same energy and angular
momentum; thus, the statistical theory gives the same quasi-stationary state.

In some experiments
\cite{btwo_dim_vortex_crystal} and simulations,
\cite{bvortex_crystals_simulation, btwodim_euler_unique_final}
it has already been found that the vortex crystal state is
sensitive to the microscopic difference in initial states
and may lead to completely different quasi-stationary states.
In the case where the quasi-stationary state is singly peaked, 
the quasi-stationary state does not depend on the microscopic details, namely,
the position of each electron in the initial state, as long as
the configuration gives
the same coarse-grained density profile,
but we still found that the quasi-stationary state 
depends on macroscopic differences in the initial state, such as the 
single- or double-ring state.

The nonergodicity of the system in the quasi-stationary state
is not an artifact due to the finite-sized mesh used in the VIC method.
As has been seen in \S\ref{ss_method},
the obtained quasi-stationary state barely depends on the mesh size as long
as it is fine enough.

On the other hand, our preliminary results suggest
that these quasi-stationary states are not really
stationary but undergo a very slow transient relaxation;
the relaxation time seems to be proportional to the total number
of particles when the time is measured in terms of global rotation
period. 
This indicates that the relaxation time required to access the ``true''
stationary state for systems of a large number of particles
such as those in our simulations and experiments
is much longer than the observation time,
even longer than the time of electron confinement in experiments,
which suggests that the true stationary state is hardly obtained
in such systems and the quasi-stationary state is realized for quite a long
time.

%We believes that the non-ergodicity of the system is not the effect
%of vortex-in-cell method. Changing the cell size of the VIC method
%gives almost no difference to the quasi stationary state, as mentioned
%in \S\ref{ss_method}, and some of our simulation results are consistent
%with experiments, such as the states close to the minimum enstrophy
%states. This indicates that the lack of ergodicity is inherent
%in this system.

As we have mentioned in \S\ref{sss_peak_density},
there is a tendency that the peak density of the initial state is
conserved.
This property, the conservation of the highest density, may be one of
the reasons that the quasi-stationary state is apparently very close to the 
minimum enstrophy state for a certain parameter region, because
the minimum enstrophy state is realized only in the parameter region
where the highest density in
the minimum enstrophy state is nearly equal to the highest density in the
initial state. To clarify this point, further study is required.

%%%%%%%%%%%%%%%%%%%%%%%%%%%%%%%%%%%%%%%%%%%%%%%%%%%%%%%%%
\section*{Acknowledgments}
We would like to thank to Professor Y. Kiwamoto, Professor M. Sakagami,
Dr. Y. Yatsuyanagi and T. Yoshida for valuable discussions and comments.

%\appendix
%\section{First Appendix} %Empty argument \section{} yields `Appendix'. 
%
%\section{Second Appendix}

\bibliographystyle{jpsj}
\bibliography{ronbun-cn-kyoto}

\end{document}